\RequirePackage{fix-cm}
\documentclass[tex,tightenlines,twocolumn,epjc3,tightenlines]{svjour3}
\smartqed  
\usepackage{graphicx}
\usepackage{graphics}
\usepackage{amsmath, amssymb}
\usepackage{graphics,bm}
\usepackage{graphicx}
\usepackage{bbold}
\usepackage{slashed}
\usepackage{feynmf}
\usepackage{wrapfig}
\usepackage[usenames]{color}     

\newcommand{\be}{\begin{equation}}
\newcommand{\ee}{\end{equation}}

\newcommand{\bqa}{\begin{eqnarray}}
\newcommand{\eqa}{\end{eqnarray}}

\def\square{\vcenter{\vbox{\hrule height.4pt
          \hbox{\vrule width.4pt height4pt
          \kern4pt\vrule width.3pt}\hrule height.4pt}}}
  
\begin{document}

\title{
Quark and pion condensates at finite isospin density in 
chiral perturbation theory}

\author{Prabal Adhikari\thanksref{e1,addr1,addr2} \and Jens O. Andersen
  \thanksref{e2,addr2}}
\thankstext{e1}{e-mail: prabal.adhikari@wellesley.edu }

\thankstext{e2}{e-mail: andersen@tf.phys.ntnu.no} 
\institute{
  Wellesley College, Department of Physics, 106 Central Street,
  Wellesley, MA 02481, United States 
  \label{addr1}
          \and
  Department of Physics, 
Norwegian University of Science and Technology, H{\o}gskoleringen 5,
N-7491 Trondheim, Norway
          \label{addr2}
}            

\maketitle
\begin{abstract}
  In this paper, we consider two-flavor QCD at zero temperature and finite
  isospin chemical potential $\mu_I$ using a model-independent analysis within
  chiral perturbation theory at next-to-leading order. We calculate the
  effective
  potential, the chiral condensate and the pion condensate in the pion-condensed
  phase at both zero and nonzero pionic source. We compare our finite pionic
  source results for the chiral condensate and the pion condensate with recent
  (2+1)-flavor lattice QCD results.
Agreement with lattice results generally 
improves as one goes from leading order to next-to-leading order.
\end{abstract}

\section{Introduction}
\label{intro}
Quantum Chromodynamics (QCD) has a rich phase structure
as a function of temperature and quark chemical
potentials~\cite{raja,alford,fukurev}.
The phases are characterized
by their symmetry and symmetry-breaking properties. The QCD vacuum
breaks chiral symmetry, a symmetry which is unbroken at the
level of the Lagrangian itself (for massless quarks).
The order parameter for
chiral symmetry breaking of the QCD vacuum is the chiral
condensate, 
\begin{equation}
\begin{split}
\langle \bar{\psi}\psi\rangle_{0}\ ,
\end{split}
\end{equation}
a zero-momentum (spatially homogeneous) state analogous to the
energetically favored Cooper pairing due to the
attractive phonon interactions in the Bardeen-Cooper-Schrieffer (BCS) theory of
superconductivity~\cite{BCS}. The analogy between chiral symmetry breaking and
Cooper-pair formation was first pointed out by Nambu and
Jona-Lasinio~\cite{Nambu}, a physical picture that is affirmed by the presence
of Goldstone modes, which are the low energy excitations around the chiral
symmetry-broken QCD vacuum. The Goldstone modes are the three pions
($\pi^{\pm}$, $\pi^{0}$) in QCD,
whose symmetries are consistent with Goldstone's
theorem~\cite{Goldstone} assuming the following symmetry breaking pattern
\begin{equation}
\begin{split}
SU(N_{f})_{L}\times SU(N_{f})_{R} \rightarrow SU(N_{f})_{V}
\end{split}
\end{equation}
for $N_{f}=2$. The symmetry group of the Lagrangian has $2(N_{f}^{2}-1)$
generators and that of the vacuum has $N_{f}^{2}-1$ generators, which leads
to exactly $N_{f}^{2}-1$ Goldstone modes.

Surprisingly, while the evidence for chiral symmetry breaking is convincing,
the chiral condensate itself is not a physical observable as is evident through
the leading-order Gell-Mann-Oakes-Renner (GOR)
relation~\cite{GOR}, valid at zero temperature and density
\begin{equation}
\begin{split}
  m_{\pi}^{2}f_{\pi}^{2}=-(m_{u}+m_{d})\langle \bar{\psi}\psi\rangle_{0}
  +\mathcal{O}(m_{q}^{2}f_{\pi}^2)\ ,\\
\end{split}
\end{equation}
where $m_{\pi}$ is the pion mass, $f_{\pi}$ is the pion decay constant, $m_{u}$
and $m_{d}$ are the up and down quarks masses respectively and $q=u,d$. The GOR
relation shows that only the product of the quark mass and the chiral condensate
can be measured indirectly through a measurement of the pion mass $m_{\pi}$ and
the pion decay constant $f_{\pi}$. Furthermore, in the chiral limit, the pion
mass is zero confirming Nambu's physical picture of chiral symmetry breaking.

The strength of chiral symmetry breaking, as measured by the magnitude of the
chiral condensate, changes depending on the physical environment. In the
presence of a magnetic field particles are largely restricted to moving in the
direction of the magnetic field, an effect known as dimensional
reduction~\cite{magcat}. This leads to the strengthening of the
quark-antiquark pairing in the chiral condensate channel, an effect analogous
to the guaranteed presence of bound states for any potential well in
one-dimensional quantum mechanics, (i.e. Cooper's theorem). The chiral
condensate for the up-quark-up-antiquark pairing is more enhanced than that
of the down-quark-down-antiquark pairing. 

On the other hand, thermal fluctuations, due to the presence of a heat bath,
have
an opposite effect on the strength of the chiral condensate. Lattice
calculations show that chiral symmetry is ``restored" at a temperature of
approximately $T_{c}^{\chi}=155\ {\rm MeV}$
~\cite{aoki,aoki2,borsa,baza0,baza}
though strictly speaking the
transition is only a crossover.
The crossover temperature is defined by the peak
of the chiral susceptibility.
This temperature is slightly less than the
crossover temperature for the deconfinement transition,
$T_{c}^{\rm decon}\approx170\ {\rm MeV}$.
However
this temperature difference is observable dependent.
In most cases, $T_{c}^{\rm decon}$
has been determined by the behavior of the Polyakov loop.
Recently, it has been defined by the behavior of the quark entropy
and in this case the two crossover temperatures agree within
errors~\cite{baza}.

A model-independent analysis within
next-to-leading order (NLO) chiral perturbation theory ($\chi$PT) shows that 
\begin{equation}
\begin{split}
  \frac{\langle\bar{\psi}\psi \rangle_{T}}{\langle\bar{\psi}\psi \rangle_{0}}=
  1-\frac{\frac{2}{3}\frac{N_{f}^{2}-1}{N_{f}}}{8f_{\pi}^{2}}T^{2}+\cdots
\end{split}
\end{equation}
with the chiral condensate decreasing quadratically with temperature ($T$)
assuming $T\ll 4\pi f_{\pi}$, the regime of validity of $\chi$PT, with the
coefficient depending on the number of flavors
($N_{f}$)~\cite{gerber,Koch,Shushpanov}.

Furthermore, the presence of matter can also have an effect on the chiral
condensate. For instance, within nucleons, the valence quarks can expel the
chiral condensate as has been shown (in a model-independent calculation~\cite{Cohen}) using
the Feynman-Hellman theorem. The physics is quite intuitive -- the
gluons that couple quarks or quarks and antiquarks, favor the formation of
protons and neutrons when the quark chemical potential is approximately a third
of the proton mass (nucleon density at saturation). As more gluons become
confined in protons and neutrons, fewer are confined within the chiral
condensate leading to its reduction.  The deviation from the vacuum value of
the chiral condensate $\langle\bar{\psi}\psi\rangle_{0}$ at low nuclear
densities $\rho_N$ is
\begin{equation}
\begin{split}
  \frac{\langle\bar{\psi}\psi \rangle_{\rho}}{\langle\bar{\psi}\psi \rangle_{0}}
  =1-\frac{\sigma_{N}}{m_{\pi}^{2}f_{\pi}^{2}}\rho_{N}+\cdots\ ,
\end{split}
\end{equation}
where $\sigma_{N}$ is the pion-nucleon sigma term and $\rho_{N}$ is the nucleon density. We note that $\sigma_{N}=59\pm7 {\rm\ MeV}$ and has been determined empirically using modern scattering data and baryon $\chi$PT at $\mathcal{O}(p^{3}$)~\cite{sigmaN}. The nucleon density at complete expulsion is
\begin{equation}
\begin{split}
  \rho_{N}^{\chi}\equiv\frac{m_{\pi}^{2}f_{\pi}^{2}}{\sigma_{N}}\sim 
  (110\ {\rm MeV})^{3}\ ,
\end{split}
\end{equation}
which is smaller than the scale $~(4\pi f_{\pi})^{3}$, well within the regime of validity of $\chi$PT~\cite{Cohen}. The uncertainty in the saturation density arises largely due to the uncertainty in determining $\sigma_{N}$~\cite{sigmaN}.
In this paper, we focus on the nature of condensates within next-to-leading
order, finite isospin
$\chi$PT, which is the effective field theory of QCD valid at energies much
lower than the typical hadronic scales, i.e.
\begin{equation}
\begin{split}
\frac{p_{\chi}}{4\pi f_{\pi}}\ll 1\ ,
\end{split}
\end{equation}
where $p_{\chi}$ is a parameter with mass dimension 1. The quantities
relevant
for this paper include momentum $p$, the isospin chemical potential
$\mu_{I}$ and a pseudoscalar, pionic source $j$~\cite{Scherer}. 

We will focus not only on the behavior of the
chiral condensate but also on the pion
condensate 
\begin{equation}
\begin{split}
\langle\pi^{\pm} \rangle_{\mu_{I}}\ .
\end{split}
\end{equation}
In the vacuum phase of QCD, i.e. for values
of $|\mu_I|\le\mu_{I}^c\equiv m_{\pi}$ it vanishes, while for
larger values of $\mu_I$, it is nonzero and 
we enter the pion-condensed phase of QCD. It is further known that pion
condensates due to their electromagnetic charge form currents in a
superconducting phase when a weak external magnetic field is
present~\cite{isomag}. For larger magnetic fields, the pion condensate 
attains a spatially inhomogeneous structure in the form of
a single vortex or a triangular vortex lattice similar in nature to the vortex
lattice in type-II superconductors~\cite{Abrikosov} explained by BCS
theory~\cite{BCS}.

Chiral perturbation theory at tree-level shows that the decrease in the size of
the chiral condensate that occurs due to the formation of pion condensates is
exactly compensated for by an increase in the pion condensate. In particular,
\begin{equation}
\begin{split}
\label{chiralcircle}
\langle \bar{\psi}\psi\rangle_{\mu_{I}}^{2}+\langle \pi^{+}\rangle_{\mu_{I}}^{2}
=\langle \bar{\psi}\psi\rangle_{0}^{2}\ .
\end{split}
\end{equation}
At low isospin chemical potentials, 
\begin{equation}
\begin{split}
\frac{\mu_{I}-m_{\pi}}{m_{\pi}}\ll 1\ ,
\end{split}
\end{equation}
the behavior of the chiral condensate in the pion-condensed phase relative to
the normal vacuum from model-independent and tree-level calculations within
$\chi$PT~\cite{son} is
\begin{equation}
\begin{split}
  \frac{\langle\bar{\psi}\psi\rangle_{\mu_{I}}}{\langle\bar{\psi}\psi\rangle_{0}}
  &=1-\frac{1}{2m_{\pi}f_{\pi}^{2}}n_{I}+\cdots\\
\end{split}
\end{equation}
where $n_{I}$ is the tree-level isospin density, which at low densities scales linearly
with the isospin chemical potential. It is worth noting that the ratio of the medium to vacuum chiral condensates (due to the expulsion of
the chiral condensate by the formation of the pion condensed phase) is
analogous in structure to the ratio found in nucleons due to expulsion of the chiral condensate through the pairing of the valence quarks $\langle qqq\rangle_{\rho}$. 

Recently there have been lattice computations of finite isospin
QCD~\cite{latiso}, which does not suffer from the fermion sign problem. This is
due to the complex phase cancellation between the up and down quarks which have
equal and opposite isospin numbers. Lattice QCD shows that the chiral structure
of Eq.~(\ref{chiralcircle}) is not preserved away from the critical isospin
chemical potential. This violation is also observed in model-dependent
calculations within the Nambu-Jona-Lasinio (NJL) model \cite{NJLisospin}.
For a recent review of meson condensation, see Ref.~\cite{mannarev}.

In this paper, we perform model-independent calculations of the chiral and
pion condensates in the pion-condensed phase at next-to-leading order within
$\chi$PT. This requires 
the effective potential $V_{\rm eff}$ at NLO
in the presence of a pionic source. This part of the calculation
turns out to be a generalization of the result
obtained in~\cite{usagain}.

The paper is organized as follows. In the next section, we briefly
discuss the chiral Lagrangian and the ground state in the presence
of a nonzero isospin chemical potential. In Sec.~\ref{eft},
we derive the effective potential at next-to-leading order in $\chi$PT
including a pionic source.
In Sec.~\ref{calculations}, we calculate the zero-temperature quark and pion
condensates at finite $\mu_I$. In Sec.~\ref{numerics}, we plot the quark and
pion condensates using lattice QCD parameters. At finite pionic source, we
compare our results with the available lattice QCD data.

\section{$\chi$PT Lagrangian}
The Lagrangian of massless two-flavor QCD
has a local $SU(3)$ gauge symmetry in addition
to the global $SU(2)_L\times SU(2)_R\times U(1)_B$ symmetries.
For nonzero quark
masses in the isospin limit, i.e for $m_u=m_d$, the symmetries
are $SU(2)_V\times U(1)_B$.
Adding a quark chemical potential $\mu_q$ for each quark,
the symmetry is $U(1)_{I_3}\times U(1)_B=U(1)_u\times U(1)_d$.
In the pion-condensed phase, the $U(1)_{I_3}$ symmetry is broken and a Goldstone boson, which is a linear combination of both the charge eigenstates ($\pi^{\pm}$), forms

Chiral perturbation theory is a low-energy effective theory for QCD based on
the symmetries and degrees of freedom~\cite{wein,gasser1,gasser2,bein}.
In two-flavor QCD, the
degrees of freedom are the pions, while for three-flavor QCD we have
additionally the charged and neutral kaons as well the eta.
In the low-energy expansion of the Lagrangian in $\chi$PT, each
covariant derivative counts as order $p$, while a 
quark mass term counts as order $p^2$.
We begin with the chiral Lagrangian in the isospin limit
at $\mathcal{O}(p^{2})$
\bqa
\mathcal{L}_{2}={f^2\over4}{\rm Tr}
  \left [\nabla_{\mu} \Sigma^{\dagger} \nabla^{\mu}\Sigma     \right ]
+{f^2\over4} {\rm Tr}
 \left [ \chi^{\dagger}\Sigma+\Sigma^{\dagger}\chi\right ]\; ,
\label{lag0}
\eqa
where $\Sigma$ parameterizes the Goldstone boson manifold
(see Eqs.~(\ref{para})--(\ref{final}) below),
$f$ is the bare pion decay constant,
\begin{align}
\chi={2B_0M}
+{2iB_0j_1}\tau_1+{2iB_0j_2}\tau_2
\;,
\end{align}
where
$M={\rm diag}(m,m)$ is the quark mass matrix
and  $-f^2B_0$ is the tree-level quark condensate. 
We have introduced a pionic source in $\chi$, which is
necessary for calculating the pion condensate.
$\tau_a$ represent the Pauli matrices
and the covariant derivatives are defined as
\bqa
\nabla_{\mu} \Sigma&\equiv&
\partial_{\mu}\Sigma-i\left [v_{\mu},\Sigma \right]\,,\\ 
\nabla_{\mu} \Sigma^{\dagger}&=&
\partial_{\mu}\Sigma^{\dagger}-i [v_{\mu},\Sigma^{\dagger} ]\;,
\eqa
with 
\bqa\nonumber
v_{\mu}&=&
\delta_{\mu0}{\rm diag}(\mu_u,\mu_d)
\\
&=&
\delta_{\mu0}{\rm diag}
(\mbox{$1\over3$}\mu_B+\mbox{$1\over2$}\mu_I,\mbox{$1\over3$}\mu_B
-\mbox{$1\over2$}\mu_I)\;,
\eqa
where 
$\mu_I=\mu_u-\mu_d$ is the isospin chemical potential and 
$\mu_B={3\over2}(\mu_u+\mu_d)$ is the baryon chemical potential. 
We also set $\mu_B=0$
for the purpose of this paper.\footnote{In the pion-condensed
  phase, physical quantities are independent of 
  $\mu_B~$~\cite{us}.}

In the two-flavor case, the ground state in 
$\chi$PT is parametrized as~\cite{son}
\bqa
\Sigma_{\alpha}&=&e^{i\alpha(\hat{\phi}_1\tau_1+\hat{\phi}_2\tau_2)}
=\mathbb{1}
\cos\alpha
+i(\hat{\phi}_1\tau_1+\hat{\phi}_2\tau_2)\sin\alpha\;,
\eqa
where $\alpha$ at tree level can be interpreted as a rotation angle
and $\hat{\phi}_1^2+\hat{\phi}_2^2=1$ to ensure the normalization of
the ground state, i.e. $\Sigma^{\dagger}_{\alpha}\Sigma_{\alpha}=\mathbb{1}$.
In the remainder of the paper, we set
$\hat{\phi_1}=1$, $\hat{\phi}_2=0$ and $j_2=0$, $j_1=j$.

The matrix $\tau_1$ generates the rotations and we can write the
rotated vacuum as $\Sigma_{\alpha}=A_{\alpha}\Sigma_0A_{\alpha}$,
where $A_{\alpha}=e^{i{\alpha\over2}\tau_1}$ and $\Sigma_0=\mathbb{1}$
is the trivial vacuum.

We also need to parametrize the fluctuations around the condensed vacuum, which
requires some care. Since the vacuum is rotated, we must also
rotate the generators of the fluctuations in the same manner.
This was discussed in Ref.~\cite{kim} and an explicit example
was given in Ref.~\cite{usagain}.
The field $\Sigma$ is written as
\bqa
\Sigma&=&L_{\alpha}\Sigma_{\alpha}R_{\alpha}^{\dagger}\;,
\label{para}
\eqa
with
\bqa
\label{la}
L_{\alpha}&=&A_{\alpha}UA_{\alpha}^{\dagger}\;,\\
R_{\alpha}&=&A_{\alpha}^{\dagger}U^{\dagger}A_{\alpha}\;.
\label{ra}
\eqa
Here $U$ is the $SU(2)$ matrix that parametrizes the fluctuations around
the ground state $\Sigma_0=\mathbb{1}$
\bqa
U&=&e^{i{\phi_a\tau_a\over2f}}\;.
\eqa
Combining Eqs.~(\ref{para})--(\ref{ra}), the expression for $\Sigma$ is
\bqa
\Sigma&=&A_{\alpha}(U\Sigma_0U)A_{\alpha}
=A_{\alpha}U^2A_{\alpha}\;,
\label{final}
\eqa
which reduces to $\Sigma=U^2$ for $\alpha=0$ as required.

In order to calculate the effective potential and the condensates
to NLO, we need to evaluate the path integral in the Gaussian approximation.
In order to do so, we must  expand
the Lagrangian ${\cal L}_2$
in the fields $\phi_a$ as
\bqa
{\cal L}_2&=&{\cal L}_2^{\rm static}
+{\cal L}_2^{\rm linear}+{\cal L}_2^{\rm quadratic}+...\;,
\eqa
where the terms we need are
\bqa
{\cal L}_2^{\rm static}&=&2f^2B_0
m_j
+{1\over2}f^2\mu_I^2\sin^2\alpha\;,\\ \nonumber
{\cal L}_2^{\rm linear}&=&
f\left(-2B_0\bar{m}_j
+\mu_I^2\sin\alpha\cos\alpha\right)\phi_1
\\ &&
+f\mu_I\sin\alpha\partial_0\phi_2\;,
\label{lini}
\\ \nonumber
  \mathcal{L}_{2}^{\rm quadratic}&=&\frac{1}{2}\partial_{\mu}\phi_{a}
  \partial^{\mu}\phi_{a}-
  {1\over2}m_a^2\phi_a^2
\\ && 
  +\mu_{I}\cos\alpha(\phi_{1}\partial_{0}\phi_{2}-\phi_{2}\partial_{0}\phi_{1})
\label{quadratic}
  \eqa
and the source-dependent masses are
\bqa
m_j&=&m\cos\alpha+j\sin\alpha\;,\\
\bar{m}_j&=&m\sin\alpha-j\cos\alpha\;,\\
m_{1}^{2}&=&2B_0m_j -\mu_{I}^{2}\cos{2}\alpha
\;, \\
m_{2}^{2}&=&2B_0m_j -\mu_{I}^{2}\cos^2\alpha
\;, \\
m_{3}^{2}&=&2B_0m_j 
+\mu_{I}^{2}\sin^{2}\alpha\;.
\eqa
The Lagrangian up to quadratic order reduces to that of
Ref. \cite{usagain} by setting $j=0$.
We get for the inverse propagator:
\bqa
D^{-1}&=&
\begin{pmatrix}
D^{-1}_{12}&0\\
0&P^{2}-m_{3}^{2}&\\
\end{pmatrix}\;,
\eqa
where $P=(p_0,p)$, $P^2=p_0^2-p^2$, and
the $2\times2$ submatrix is given by
\begin{align}
D^{-1}_{12}&=
\begin{pmatrix}
P^{2}-m_{1}^{2}&ip_{0}m_{12}\\
-ip_{0}m_{12}&P^{2}-m_{2}^{2}\\
\end{pmatrix}\;.
\end{align}
Here the off-diagonal mass is defined as 
\bqa
m_{12}&=&2\mu_{I}\cos\alpha\;.
\eqa
At next-to-leading order in the low-energy expansion, there are ten
different operators
in the Lagrangian~\cite{gasser1}. 
The terms relevant for the present calculations are~\cite{gerber}
\bqa
\nonumber
{\cal L}_4&=&  {1\over4}l_1\left({\rm Tr}
\left[\nabla{\mu}\Sigma^{\dagger}\nabla^{\mu}\Sigma\right]\right)^2
\\ \nonumber&&
+{1\over4}l_2{\rm Tr}\left[\nabla_{\mu}\Sigma^{\dagger}\nabla_{\nu}\Sigma\right]
    {\rm Tr}\left[\nabla^{\mu}\Sigma^{\dagger}\nabla^{\nu}\Sigma\right]
\\ && \nonumber
+{1\over16}(l_3+l_4)({\rm Tr}[\chi^{\dagger}\Sigma+\Sigma^{\dagger}\chi])^2
\\&& \nonumber
    +{1\over8}l_4{\rm Tr}\left[\nabla_{\mu}\Sigma^{\dagger}\nabla^{\mu}\Sigma\right]
{\rm Tr}[\chi^{\dagger}\Sigma+\Sigma^{\dagger}\chi]
\\ &&
+{1\over2}h_1{\rm Tr}[\chi^{\dagger}\chi]   \;.
\label{lag}
\eqa
Here $l_i$ and $h_i$ are bare couplings. The relations between the
bare and renormalized couplings $l_i^r(\Lambda)$ and
$h_i^r(\Lambda)$ are~\cite{gasser2} 
\bqa
l_i&=&l_i^r(\Lambda)-{\gamma_i\Lambda^{-2\epsilon}\over2(4\pi)^2}\left[
{1\over\epsilon}+1\right]\;,
\label{renorm1}
\\
h_i&=&h_i^r(\Lambda)-{\delta_i\Lambda^{-2\epsilon}\over2(4\pi)^2}\left[
  {1\over\epsilon}+1\right]\;,
\label{renorm2}
\eqa
where $\Lambda$ is the renormalization scale in the modified minimal
subtraction ($\overline{\rm MS}$) scheme.
The constants
$\gamma_i$ and $\delta_i$ are~\cite{gasser2}
\begin{align}
  &  \gamma_{1}=\frac{1}{3}\;,& \gamma_{2}&=\frac{2}{3}\;,& \gamma_{3}&=-{1\over2}
                                                                        \;,
\\&  \gamma_{4}=2\;,&
    \delta_{1}&=0\;.
\end{align}
Taking the derivative of Eqs.~(\ref{renorm1})--(\ref{renorm2})
with respect to $\Lambda$
and using that the bare couplings are independent of the scale,
one finds that the running couplings satisfy the equations,
\bqa
\Lambda{d\over d\Lambda}l_i^r=-{\gamma_i\over(4\pi)^2}\;,
\hspace{1cm}
\Lambda{d\over d\Lambda}h_i^r=-{\delta_i\over(4\pi)^2}\;.
\label{dh}
\eqa
These equations can be easily solved for the running couplings $l_i^r$
and $h_i^r$,
The relations between the running couplings and the
so-called low-energy constants $\bar{l}_i$
and $\bar{h}_i$ in two-flavor $\chi$PT are
\bqa
\label{2fLEC0}
l_i^r(\Lambda)&=&{\gamma_i\over2(4\pi)^2}
\left[\bar{l}_i+\log{M^2\over\Lambda^2}\right]\;,
\\
h_i^r(\Lambda)&=&{\delta_i\over2(4\pi)^2}
\left[\bar{h}_i+\log{M^2\over\Lambda^2}\right]\;.
\label{2fLEC}
\eqa
Up to a prefactor, the low-energy constants are
the running couplings evaluated at the scale
$\Lambda^2=M^2$. 
We return to this in the Sec.~\ref{numerics}.
Note that, due to $\delta_1=0$, Eq.~(\ref{2fLEC}) does not apply and
Eq.~(\ref{dh}) shows that $h_1^r(\Lambda)$ does not run.
Moreover, in the original paper~\cite{gasser1}, the authors used another set
of invariant operators than the ones (partially) listed in Eq.~(\ref{lag}).
Using the equations of motion one can obtain one from the other.
This implies relations among couplings, 
$h_1=\tilde{h}_1-\tilde{l}_4$, where $\tilde{l}_i$ and $\tilde{h}_i$
refer to the original couplings from Ref.~\cite{gasser1}. The corresponding values of
$\tilde{\gamma}_i$ and $\tilde{\delta}_i$ are the same as above, except
$\tilde{\delta}_1=2$ implying that $\tilde{h}_1$ runs.

\section{Effective potential}  
\label{eft}
Since the terms ${\cal L}_2^{\rm static}$,
${\cal L}_2^{\rm linear}$, and ${\cal L}_2^{\rm quadratic}$
as well as ${\cal L}_4^{\rm static}$ (see below) can be obtained
from the results in Ref.~\cite{usagain} by using source-dependent
mass parameters, the calculation of the effective potential here
is a straightforward generalization of the calculation therein.
However, for completeness, we include the details here.
At tree level, the effective potential $V_0$
is given by $-{\cal L}_2^{\rm static}$,
\bqa
V_0&=&-2f^2B_0
m_j
-{1\over2}f^2\mu_I^2\sin^2\alpha\;.
\label{v0} 
\eqa
The value of $\alpha$ that minimizes the
tree-level potential $V_0$ is given by
${\partial V_0\over\partial\alpha}=0$ or
$2B_0\bar{m}_j-\mu_I^2\sin\alpha\cos\alpha=0$.
The linear term ${\cal L}_2^{\rm linear}$ in Eq.~(\ref{lini})
then vanishes at the minimum of the
tree-level potential, as required.
(The surface term, $f\mu_I\sin\alpha\partial_0\phi_2$, can be
ignored).
At next-to-leading order, there are two contributions to the
effective potential, namely the static term
$V_1^{\rm static}=-{\cal L}_4^{\rm static}$
and the one-loop contribution $V_1$
arising from the Gaussian path integral involving the
  quadratic terms in the Lagrangian, ${\cal L}_2$,
  given by Eq.~(\ref{quadratic}).

The static part of the NLO effective potential is
\bqa\nonumber
V_1^{\rm static}&=&
-(l_1+l_2)\mu_I^4\sin^4\alpha
-2l_4B_0m_j\mu_I^2\sin^2\alpha
\\ && 
-4(l_3+l_4)B_0^2m_j^2
-4h_1B_0^2\left[m_j^2+\bar{m}_j^2\right]
\;,
\label{v1stat}
\eqa
which acts as counterterms in the NLO calculation.
After performing the Gaussian integral to obtain
  the one-loop correction, $V_1$, to the effective potential, we Wick
rotate to Euclidean space.
The one-loop contribution to the effective potential in Euclidean space
of a free massive boson is given by
\bqa
V_1&=&{1\over2}\int_P\log\left[P^2+m^2\right]\;,
\label{logint}
\eqa
where now $P^2=p_0^2+p^2$ and the integral is defined as
\bqa
\int_P&=&\int{dp_0\over2\pi}\int_p
=\int{dp_0\over2\pi}
\left (\frac{e^{\gamma_{E}}\Lambda^{2}}{4\pi} \right )^{\epsilon}
\int
\frac{d^{d}p}{(2\pi)^{d}}\;.
\eqa
We use dimensional regularization to regulate ultraviolet divergences
with  the momentum  integral
generalized  to $d=3-2\epsilon$ dimensions. Then the integral in
Eq.~(\ref{logint}) becomes
\bqa\nonumber
\int_P\log[P^2+m^2]&=&\int_p\sqrt{p^2+m^2}
\\ \nonumber
&=&-{m^4\over2(4\pi)^2}
\left({\Lambda^2\over m^2}\right)^{\epsilon}
\left[{1\over\epsilon}+{3\over2}+{\cal O}(\epsilon)\right]\;.
\\ && 
\label{sumint0}
\eqa
The contribution from $\pi^0$ 
can be calculated analytically in dimensional regularization
using Eq.~(\ref{sumint0}),
\bqa
V_{1,\pi^0}&=&{1\over2}\int_P\log\left[P^2+m_3^2\right]\;.
\eqa
The contribution from the charged pions requires a little more work.
Using Eq.~(\ref{sumint0}), we obtain
\bqa\nonumber
V_{1,\pi^+}+V_{1,\pi^-}&=&
{1\over2}\int_P\log[(p_0^2+E_{\pi^+}^2)(p_0^2+E_{\pi^-}^2)]
\\ &=&
{1\over2}\int_p\left[{E_{\pi^+}}+{E_{\pi^-}}\right]\;,
\eqa
where the energies $E_{\pi^{\pm}}$ are found by calculating the zeros of the
inverse propagator $D_{12}^{-1}$ and read
\bqa
\nonumber
E_{\pi^{\pm}}^{2}&=&p^2+{1\over2}\left(m_{1}^{2}+m_{2}^{2}+m_{12}^{2}\right)
\\ 
&&
\hspace{-0.5cm}
\pm{1\over2}\sqrt{4p^{2}m_{12}^{2}+(m_{1}^{2}+m_{2}^{2}
  +m_{12}^{2})^2-4m_{1}^{2}m_{2}^{2}}\;.
\label{pipo}
\eqa
In order to eliminate the divergences, 
their dispersion relations are expanded in powers of $1/p$ as
\bqa\nonumber
E_{\pi^+}+E_{\pi^-}&=&
2p+\frac{2(m_{1}^{2}+m_{2}^{2})+m_{12}^{2}}{4p}
\\ &&
\hspace{-1.5cm}
-\frac{8(m_{1}^{4}+m_{2}^{4})+4(m_{1}^{2}+m_{2}^{2})m_{12}^{2}
  +m_{12}^{4}}{64p^{3}}+...
\label{expandp}
\eqa
To this order, the large-$p$ behavior in Eq.~(\ref{expandp}) is the same
as the sum $E_1+E_2$, where the energies and masses are
$E_{1,2}=\sqrt{p^2+m_{1,2}^2+\mbox{$1\over4$}m_{12}^2}=\sqrt{p^2+\tilde{m}^2_{1,2}}$,
$\tilde{m}_1^2=m_3^2$ and $\tilde{m}_2^2=2B_0m_j$.
We can then write
\bqa
V_{1,\pi^+}+V_{1,\pi^-}
&=&V_{{\rm 1},\pi^{+}}^{\rm div}+V_{{\rm 1},\pi^{-}}^{\rm div}
+
V_{{\rm 1},\pi^{+}}^{\rm fin}+V_{{\rm 1},\pi^{-}}^{\rm fin}\;,
\eqa
where
\bqa
\label{divint}
V_{{\rm 1},\pi^{+}}^{\rm div}+V_{{\rm 1},\pi^{-}}^{\rm div}
&=&{1\over2}\int_p\left[E_1+E_2\right]
\;,\\
V_{{\rm 1},\pi^{+}}^{\rm fin}+V_{{\rm 1},\pi^{-}}^{\rm fin}
&=&\frac{1}{2}\int_{p}\left [E_{\pi^+}+E_{\pi^-}-E_1-E_2\right ]\;.
\label{subt}
\eqa
The divergent integrals in Eq.~(\ref{divint})
can be done analytically in
dimensional regularization. The
subtraction integral (\ref{subt}) is finite
and can be computed numerically.

Using Eq.~(\ref{sumint0}), the divergent
part of the one-loop contribution can be written as 
\bqa\nonumber
V_{\rm 1}^{\rm div}
&=&V_{1,\pi^0}+
V_{{\rm 1},\pi^{+}}^{\rm div}+V_{{\rm 1},\pi^{-}}^{\rm div}\\ \nonumber
&=&
-\frac{\tilde{m}_1^4}{4(4\pi)^{2}}\left [\frac{1}{\epsilon}+\frac{3}{2} +
  \log\left (\frac{\Lambda^{2}}{{\tilde{m}_1^2}}
  \right ) \right ]
\\ &&\nonumber-\frac{\tilde{m}_2^4}{4(4\pi)^{2}}
\left [\frac{1}{\epsilon}+\frac{3}{2} +
  \log\left (\frac{\Lambda^{2}}{{\tilde{m}_2^2}}
  \right ) \right ]
\\ &&
-\frac{m_3^4}{4(4\pi)^{2}}\left [\frac{1}{\epsilon}+\frac{3}{2} +
  \log\left (\frac{\Lambda^{2}}{m_{3}^{2}} \right )\right ]
\label{divpi}
\;.
\eqa
Renormalization is now carried out by adding
  Eqs.~(\ref{v0}), (\ref{v1stat}), and (\ref{divpi}),
  using Eqs.~(\ref{renorm1})--(\ref{renorm2}).
Using Eq.~(\ref{2fLEC0}), 
the renormalized effective potential is 
\bqa\nonumber
V_{\rm eff}&=&-2f^2B_0m_j-{1\over2}f^2\mu_I^2\sin^2\alpha \\
\nonumber
&&-\frac{1}{(4\pi)^{2}}\left[\frac{3}{2}-\bar{l}_{3}+4\bar{l}_{4}
  +\log\left(\frac{M^2}{\tilde{m}_{2}^{2}}\right)\right.\\ \nonumber
&&\left.+2\log\left(\frac{M^2}{m_{3}^{2}}\right)\right ]B_{0}^{2}m_{j}^{2}
\\\nonumber
&&-\frac{1}{(4\pi)^{2}}\left[{1\over2}+\bar{l}_{4}
  +\log\left(\frac{M^2}{m_{3}^{2}}\right) \right]
2B_{0}m_{j}\mu_{I}^{2}\sin^{2}\alpha\\ \nonumber
&&-\frac{1}{2(4\pi)^{2}}\left[\frac{1}{2}+\frac{1}{3}\bar{l}_{1}+\frac{2}{3}
  \bar{l}_{2}+\log\left(\frac{M^2}{m_{3}^{2}}\right) \right ]\mu_{I}^{4}
\sin^{4}\alpha\\ 
&&
-{4\over(4\pi)^2}\bar{h}_1B_0^2\left[m_j^2+\bar{m}_j^2\right]
+V^{\rm fin}_{1,\pi^{+}}+V^{\rm fin}_{1,\pi^{-}}\;.
\label{renpi}
\eqa
For zero pionic source, $j=0$, Eq.~(\ref{renpi})
reduces to the result of Ref.~\cite{usagain} after subtracting the constant term
proportional to $\bar{h}_{1}$. We note that since $h_{1}^{r}$ does not run due
to Eq.~(\ref{dh}), we have defined
$\bar{h}_{1}=(4\pi)^{2} h_{1}^{r}=(4\pi)^{2}h_{1}$.

\section{Quark and pion condensates}
\label{calculations}
In Refs.~\cite{usagain,us}, we studied the thermodynamic properties
of the pion-condensed phase of QCD at $T=0$
at next-to-leading order by calculating the
first quantum correction to the tree-level potential.
It was shown that the transition from the
vacuum phase to a pion-condensed phase is second order and takes
place at a critical isospin chemical potential $\mu_I^c=m_{\pi}$, where
$m_{\pi}$ is the physical pion mass. We continue the study of the
pion-condensed phase by calculating the quark and pion condensates.

In the isospin limit, the quark condensates
$\langle\bar{u}u\rangle$ and $\langle\bar{d}d\rangle$ are equal
and in the following we denote each of them by $\langle\bar{\psi}\psi\rangle$.
The quark and pion condensates at finite isospin are then defined
as~\footnote{Note that in the finite isospin lattice QCD simulation of Ref.~\cite{latiso},
  $\langle\bar{\psi}\psi\rangle=\langle\bar{u}u\rangle+\langle\bar{d}d\rangle$
  but in our notation
  $\langle\bar{\psi}\psi\rangle=\langle\bar{u}u\rangle=\langle\bar{d}d\rangle$.
  Consequently, there is an explicit factor of $\frac{1}{2}$ in our definition
  of $\langle\bar{\psi}\psi\rangle$. Additionally, compared to
  Ref.~\cite{latiso}, we define the pion condensate with an extra factor of
  $\frac{1}{2}$. The pionic source $\lambda$ in Ref.~\cite{latiso} corresponds
  exactly to $j$ in this paper.}
\bqa
\langle\bar{\psi}\psi\rangle_{\mu_{I}}
={1\over2}{\partial V_{\rm eff}\over\partial m}\;,
\hspace{1cm}
\langle\pi^+\rangle_{\mu_{I}}={1\over2}{\partial V_{\rm eff}\over\partial j}\;.
\eqa
At tree level, the condensates are given by the partial derivatives
of $V_0$, which yields
\bqa
\label{q1}
\langle\bar{\psi}\psi\rangle^{\rm tree}_{\mu_{I}}
&=&-f^2B_0\cos\alpha
=\langle\bar{\psi}\psi\rangle_0^{\rm tree}\cos\alpha
\;,\\
\langle\pi^+\rangle^{\rm tree}_{\mu_{I}}&=&-f^2B_0\sin\alpha
=\langle\bar{\psi}\psi\rangle_0^{\rm tree}\sin\alpha
\;,
\label{p1}
\eqa
where $\langle\bar{\psi}\psi\rangle_0^{\rm tree}=-f^2B_0$ denotes
the quark condensate in the vacuum phase. Eqs.~(\ref{q1})--(\ref{p1})
show that we can interpret $\alpha$ as a rotation angle such that
the quark condensate is rotated into a pion condensate.
As we shall see below, this interpretation is not valid at next-to-leading
order and is not seen on the lattice.
At next-to-leading order in the low-energy expansion,
the quark condensate is
\bqa\nonumber
\langle\bar{\psi}\psi\rangle_{\mu_{I}}&=&
-{f}^2{B}_0
\cos\alpha\left[1+  {1\over(4\pi)^2}\bigg(
  -\bar{l}_3+    4\bar{l}_4
 \right.\\ && \nonumber\left.
   +  \log{M^2\over\tilde{m}_2^2}
   +2\log{M^2\over m_3^2}
       \bigg){B_0m_j\over f^2}
\right.\\ &&\nonumber \left.
  +{1\over(4\pi)^2}\left(\bar{l}_4+\log{M^2\over m_3^2}\right)
  {\mu_I^2\sin^2\alpha\over f^2}
\right]
\\ &&
-{4\over(4\pi)^2}\bar{h}_1{B}_0^2{m}
+{1\over2}{\partial V_{\rm 1,\pi^+}^{\rm fin}\over\partial m}
+{1\over2}{\partial V_{\rm 1,\pi^-}^{\rm fin}\over\partial m}\;.
\label{light}
\eqa
In the limit of vanishing source $j$ and $\alpha=0$,
Eq.~(\ref{light}) is independent of the isospin chemical
potential and are consistent with expressions given in
Refs.~\cite{gasser1,gasser2}.

At next-to-leading order in the low-energy expansion, the pion condensate is
\bqa\nonumber
\langle\pi^+\rangle_{\mu_{I}}&=&
-{f}^2{B}_0
\sin\alpha\left[1+  {1\over(4\pi)^2}
  \bigg(-\bar{l}_3+4\bar{l}_4
  \right.\\ &&\left.\nonumber
    +\log{M^2\over\tilde{m}_2^2}
    +2\log{M^2\over m_3^2} 
  \bigg){B_0m_j\over f^2}
\right.\\ &&\nonumber \left.
  +{1\over(4\pi)^2}\left(\bar{l}_4+\log{M^2\over m_3^2}\right)
    {\mu_I^2\sin^2\alpha\over f^2}
  \right]
\\  && 
-{4\over(4\pi)^2}\bar{h}_1{B}_0^2{j}
+{1\over2}
{\partial V_{\rm 1,\pi^+}^{\rm fin}\over\partial j}
+{1\over2}{\partial V_{\rm 1,\pi^-}^{\rm fin}\over\partial j}\;.
\label{pioncon}
\eqa
We note that the pion condensate vanishes in the normal vacuum since $\alpha=0$ and such a vacuum only exists if the pion source is zero. However, in the presence of a pionic source, i.e. $j\neq 0$, the pion condensate is non-zero not only due to $\alpha$-dependent contributions but also a term proportional to $j$, which is independent of $\alpha$. The term arises due to the non-dynamical contribution $\tfrac{{1}}{2}h_1{\rm Tr}[\chi^{\dagger}\chi]$ in the $\mathcal{O}(p^{4})$ $\chi$PT Lagrangian.
\section{Results and discussion}
\label{numerics}
In this section, we present our numerical results for the chiral condensate
and the pion condensate both at zero and non-zero
pionic source. We compare the non-zero pionic source results with lattice
simulations for which lattice data are available.
Finite isospin QCD on the lattice is studied by adding an
explicit pionic source since spontaneous symmetry breaking in finite volume is
forbidden. Obtaining the chiral and pion condensate then requires not just
taking the continuum limit but also extrapolating to a zero external source,
which is technically challenging on the lattice
 
The quark condensate is given by Eqs.~(\ref{light}), while
the pion condensate is given by
Eq.~(\ref{pioncon}). The value of $\alpha$ in the equations is
found by extremizing the effective potential, i.e. solving
${\partial V_{\rm eff}\over\partial\alpha}=0$.

\subsection{Definitions and choice of parameters}
The chiral condensate depends on the low-energy constant $\bar{h}_{1}$ of
two-flavor $\chi$PT, which is unphysical and undeterminable within
$\chi$PT~\cite{Hr2ref1,Hr2ref2}. Furthermore, $\bar{h}_{1}$ is scale-independent
and does not affect the ground state value of $\alpha$. 
Consequently, we define the quark and pion condensate deviations relative to the
values of the respective condensates at zero isospin and zero pionic source.
The definitions of the condensate deviations~\footnote{Note that compared to
  Ref.~\cite{latiso}, our definitions of the condensate deviations carry an
  explicit factor $2$, which is exactly compensated by the difference of a
  factor of $\frac{1}{2}$ each in our definitions of
  $\langle\bar{\psi}\psi\rangle$ and $\langle\pi^{+}\rangle$.} are~\cite{gergy3}
\begin{align}
\label{deviation1}
\Sigma_{\bar{\psi}\psi}&=-\frac{2m}{m_{\pi}^{2}f_{\pi}^{2}}\left[\langle\bar{\psi}
    \psi\rangle_{\mu_{I}}-\langle\bar{\psi}\psi\rangle_{0}^{j=0}
\right]+1\;,\\
\Sigma_{\pi}&=-\frac{2m}{m_{\pi}^{2}f_{\pi}^{2}}\langle\pi^{+}\rangle_{\mu_{I}}\;,
    \label{deviation2}
\end{align}
where $m$ is the degenerate mass of the up and down quarks,
$m_{\pi}$ is the pion mass, and $f_{\pi}$ is the pion decay constant.
$\langle O\rangle_{\mu_{I}}$ is the value of the condensate $O$ at an isospin
chemical potential $\mu_{I}$ and a pionic source $j$.
$\langle \bar{\psi}\psi\rangle_{0}^{j=0}$ is the value of the chiral condensate
when $\mu_{I}=0$ and $j=0$. The definition of the chiral condensate
deviation, $\Sigma_{\bar{\psi}\psi}$, ensures that it is equal to $1$ when
$\mu_{I}=0$ and $j=0$ and the definition of the pion condensate deviation does
not contain a trivial subtraction of the pion condensate at zero pionic source
and zero isospin, $\langle\pi^{+}\rangle_{0}^{j=0}$, since it equals zero. Since the pion condensate deviation, $\Sigma_{\pi}$, is a rescaled, dimensionless quantity proportional to the pion condensate of Eq.~(\ref{pioncon}), it is worth noting again that just like the pion condensate it vanishes in the absence of a pionic source if $\alpha=0$ but when a pionic source is turned on, the deviation becomes non-zero not only due to $\alpha$-independent terms but also a term proportional to $j$ that is independent of $\alpha$.
Furthermore, the definitions of the deviations ensure that the following constraint is satisfied
at tree level including for any pionic source $j$
\begin{equation}
\begin{split}
  \big(\Sigma_{\bar{\psi}\psi}^{\rm tree}\big)^{2}
  +\big(\Sigma_{\pi}^{\rm tree}\big)^{2}=1\ ,
\end{split}
\end{equation}
which is consistent with Eqs.~(\ref{q1}) and (\ref{p1}).
However the constraint is not satisfied at next-to-leading order as will be
evident. 

For our calculation of the condensate deviations, we choose the
following values of the quark masses~\cite{BMW}
\begin{align}
m_{u}&=2.15 \;\text{MeV}\;,
m_{d}=4.79\;\text{MeV}\;,
\\ 
m&=\frac{m_{u}+m_{d}}{2}=3.47\;\text{MeV}\;.
\end{align}
Since we want to compare our results 
to those of recent lattice calculations~\cite{private},
we choose their values for the pion mass and the pion decay constant,
\bqa
\label{masses}  
m_{\pi}&=&131\pm3\;{\rm MeV},\;
f_{\pi}={128\pm3\over\sqrt{2}}{\rm MeV}\;.
\label{decays}
\eqa
It is important to point out that the quark masses quoted above 
from~\cite{BMW} are
not the quark masses of~\cite{private} (they are not known).
The quark masses of~\cite{BMW} correspond to a pion mass of approximately
135 MeV, i.e. approximately 3\% higher than the one used in the lattice simulations. In order to improve the overall confidence in our comparison, we
therefore vary the quark mass $m$ by 5\%, which is consistent with 
uncertainty quoted.

The LECs of two-flavor $\chi$PT and their respective uncertainties are defined
at the scale $\Lambda^2=2B_0m$ through 
Eq.~(\ref{2fLEC})~\cite{cola}~\footnote{Note that we take the derivative of 
$V_{\rm eff}$ w.r.t. the quark mass, $m$, before we choose the scale $\Lambda^2=2B_0m$.}
\begin{align}
  \label{variasjon1}
\bar{l}_{1}&=-0.4\pm 0.6\;,
&\bar{l}_{2}&=4.3\pm 0.1\;,\\
\bar{l}_{3}&=2.9 \pm 2.4\;,
&\bar{l}_{4}&=4.4 \pm 0.2\;\\
\bar{h}_{1}&=-1.5\pm 0.2\;.&
\label{variasjon2}
\end{align}
The LEC, $\bar{h}_{1}$, was deduced using the value of $H^{r}_{2}$ and its uncertainties in Ref.~\cite{Hr2ref1} and the mapping of 3-flavor LECs to 2-flavor LECs discussed in Ref.~\cite{gasser2}.
The chiral condensate deviation is independent of $\bar{h}_{1}$ at all values of $j$ but the pion condensate deviation depends on $\bar{h}_{1}$ at finite $j$. There is at least another choice of $\bar{h}_{1}$ in 
literature~\cite{gasser1}, which happens to be model-dependent
(calculations based on $\rho$-dominance). 
However, the pion condensate deviation is not affected significantly by this 
choice~\cite{martin}.
The physical pion mass $m_{\pi}$ and the physical pion decay constant
$f_{\pi}$ can be calculated within $\chi$PT at NLO~\cite{gasser1},
\bqa
m_{\pi}^2&=&2B_0m\left[1-{B_0m\over(4\pi)^2f^2}\bar{l}_3\right]\;,
\label{mpi}
\\ 
f_{\pi}^2
&=&f^2\left[1+{4B_0m\over(4\pi)^2f^2}\bar{l}_4\right]\;.
\label{fpi}
\eqa
Given the values $m_{\pi}$, $f_{\pi}$, $\bar{l}_3$, and  $\bar{l}_4$,
we can calculate the parameters $f$ and $2B_0m$ appearing in the
chiral Lagrangian:
\begin{align}
\label{pp1}
  m_{\pi,0}^{\rm cen}&=132.49
                       \,{\rm MeV}\;,
  &f^{\rm cen}=84.93
    \,{\rm MeV}\;,\\
  m_{\pi,0}^{\rm min}&=128.24
                       \,{\rm MeV}\;,
  &f^{\rm min}=83.29
    \,{\rm MeV}\;, \\
  m_{\pi,0}^{\rm max}&=136.91
                       \,{\rm MeV}\;,
  &f^{\rm max}=86.54
    \,{\rm MeV}\;,
\label{pp3}
\end{align}
where $m_{\pi,0}^{2}\equiv 2B_{0}m$. Using this relation, we can calculate $B_{0}$,
which also depends on the tree-level pion mass and the continuum value of the
quark mass. 

\subsection{$\alpha_{\rm gs}$}
\begin{figure}[htb]
\centering
\includegraphics[width=0.4\textwidth]{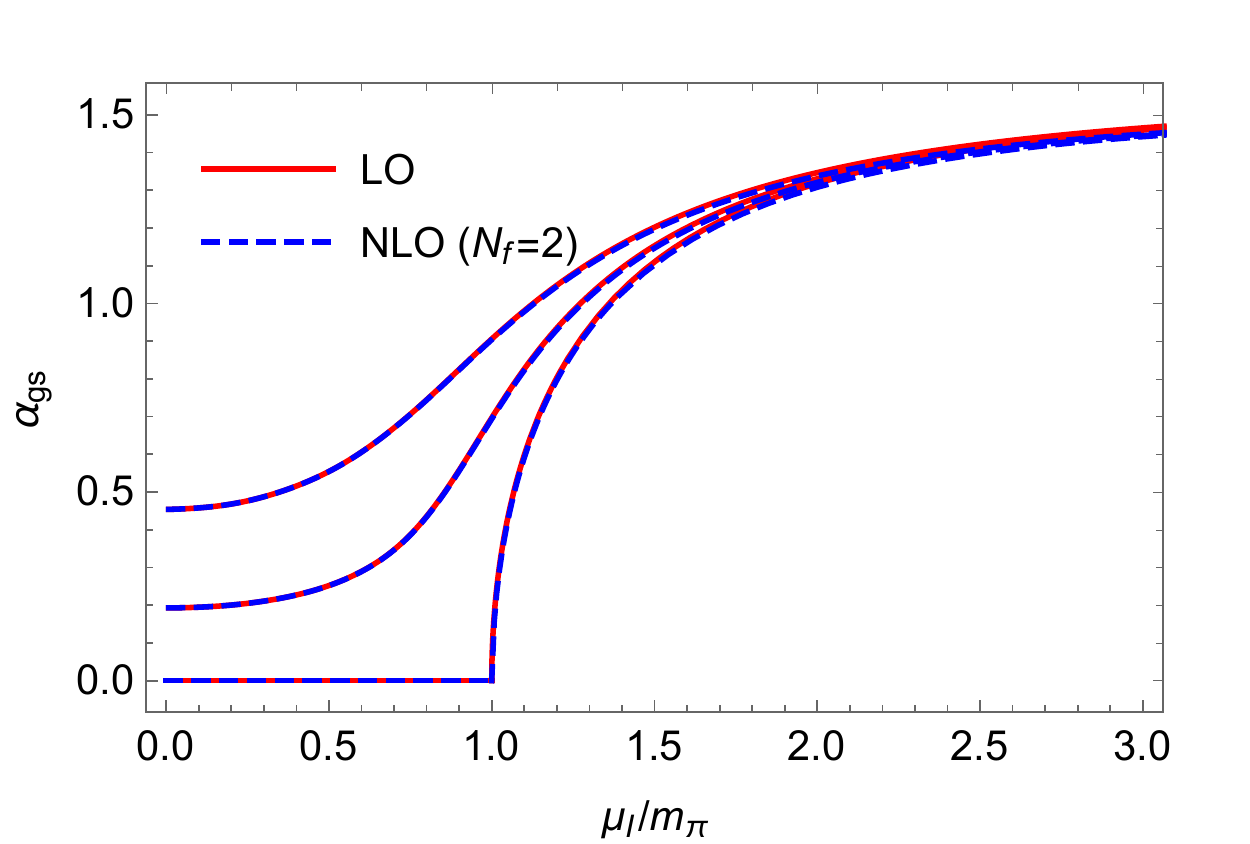}
\caption{$\alpha_{\rm gs}$ as a function of $\mu_I/m_{\pi}$ 
from below for $j=0$,
  $j=0.00517054m_{\pi}$ and
  $j= 0.0129263 m_{\pi}$. Red solid lines are LO results
  and blue dashed lines are NLO results.}
\label{alphags}  
\end{figure}

Before we discuss the condensates, we present the solution $\alpha_{\rm gs}$
to the equation ${\partial V_{\rm eff}\over\partial\alpha}=0$ as function
of $\mu_I/m_{\pi}$ for three different values of the source $j$.
The resulting curves are shown in Fig.~\ref{alphags}.
The red solid lines are the leading-order results, while 
the dashed blue lines are the next-to-leading order results.
In all three cases, the difference is very small.

For $j=0$ the  curve for $\alpha_{\rm gs}$ is not smooth at $\mu_I=m_{\pi}$,
which simply reflects the second-order transition from the vacuum phase
to the pion-condensed phase. For nonzero source, the isospin symmetry is
explicitly broken resulting in nonzero values of 
$\alpha_{\rm gs}$ for all values of $\mu_I$. Moreover, the curves
are smooth, which is due to the cross-over nature of the transition, rather than
a second-order phase transition. In the limit $\mu_I\rightarrow\infty$, the curves
approach the asymptotic value of $\alpha_{\rm gs}={\pi\over2}$.

\subsection{Condensates at $j=0$}
\label{sectionj0}

\begin{figure}[htb]
\centering
\includegraphics[width=0.45\textwidth]{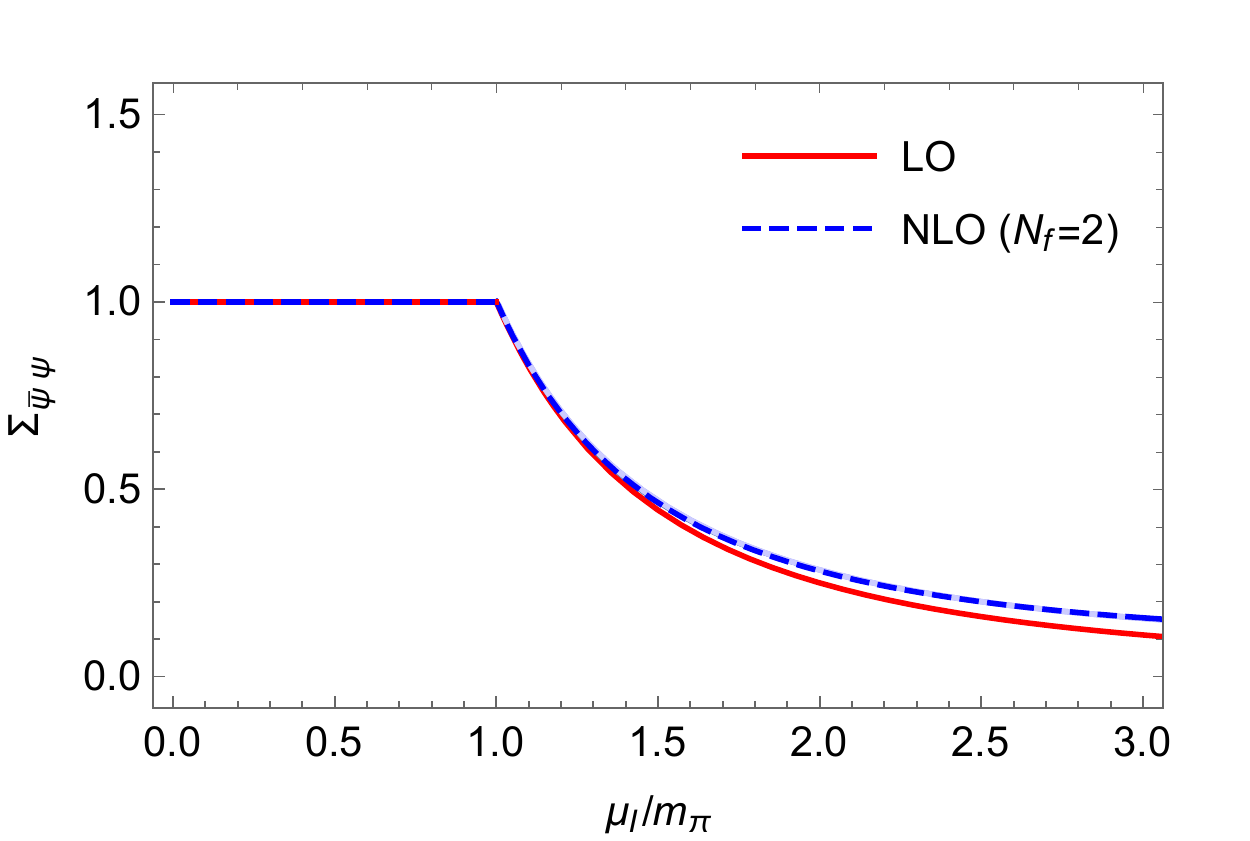}\,\,\,\,\,
\includegraphics[width=0.45\textwidth]{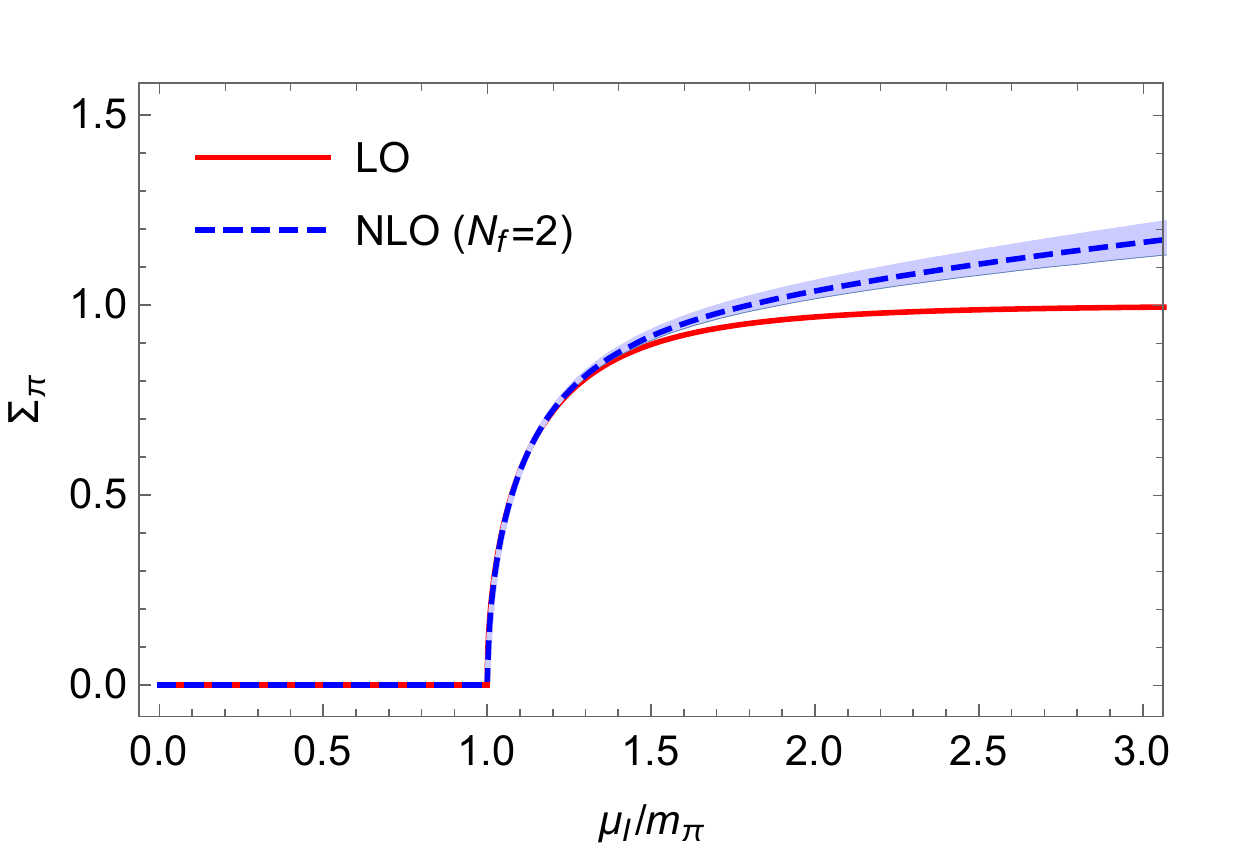}\,\,\,\,\,
\caption{Top: Quark condensate deviation (normalized to $1$) from the normal
  vacuum value, $\Sigma_{\bar{\psi}\psi}$, at $T=0$. Bottom: Pion condensate
  deviation from the normal vacuum value (which is 0), $\Sigma_{\pi}$, at $T=0$
  and $j=0$.
  See text in Section~\ref{sectionj0}
  for details.}
\label{fig:ccpc}       
\end{figure}
In Fig.~\ref{fig:ccpc}, we plot the $T=0$ quark condensate deviation
(which is normalized to 1) and the pion condensate deviation defined in
Eqs.~(\ref{deviation1}) and (\ref{deviation2}) respectively. In the upper panel
of Fig.~\ref{fig:ccpc}, we plot the
tree level chiral condensate deviation in solid red and the next-to-leading
order deviation in dashed blue. Similarly, in
the bottom panel of Fig.~\ref{fig:ccpc}, we plot the tree-level
pion condensate deviation in solid red and the 
the next-to-leading order deviation in dashed blue. Note that at $j=0$, the tree-level deviations are independent of the quark mass.
The light blue shaded regions in the two
panels of
Fig.~\ref{fig:ccpc} represent the uncertainty in the condensate deviations due
to the uncertainty in the values
of the pion mass and the pion decay constant from the lattice, the uncertainty
in
the LECs, which arises due to experimental uncertainties, and the uncertainty in the lattice quark masses at the $5\%$ level which is consistent with results in Ref.~\cite{BMW}. We note that the
uncertainty in the condensate deviations is dominated by the uncertainties in
the pion mass and pion decay constant with the uncertainties in the LECs and the quark masses not
contributing significantly.

We find that relative to the tree-level condensate deviations, the
next-to-leading condensate deviations are moderately larger
  for the chiral condensate and significantly larger for the pion condensate.
The magnitude of the chiral deviation condensate at
next-to-leading order decreases more slowly and the magnitude of the pion
condensate increases more rapidly compared to their
respective tree-level values. Furthermore, the tree-level pion condensate
deviation asymptotes to $1$ very efficiently, a behavior which is absent at
next-to-leading order.

\subsection{Condensates at finite $j$ and comparison with lattice QCD}

\label{latcomp}
\begin{figure}[t!]
\centering
\includegraphics[width=0.45\textwidth]{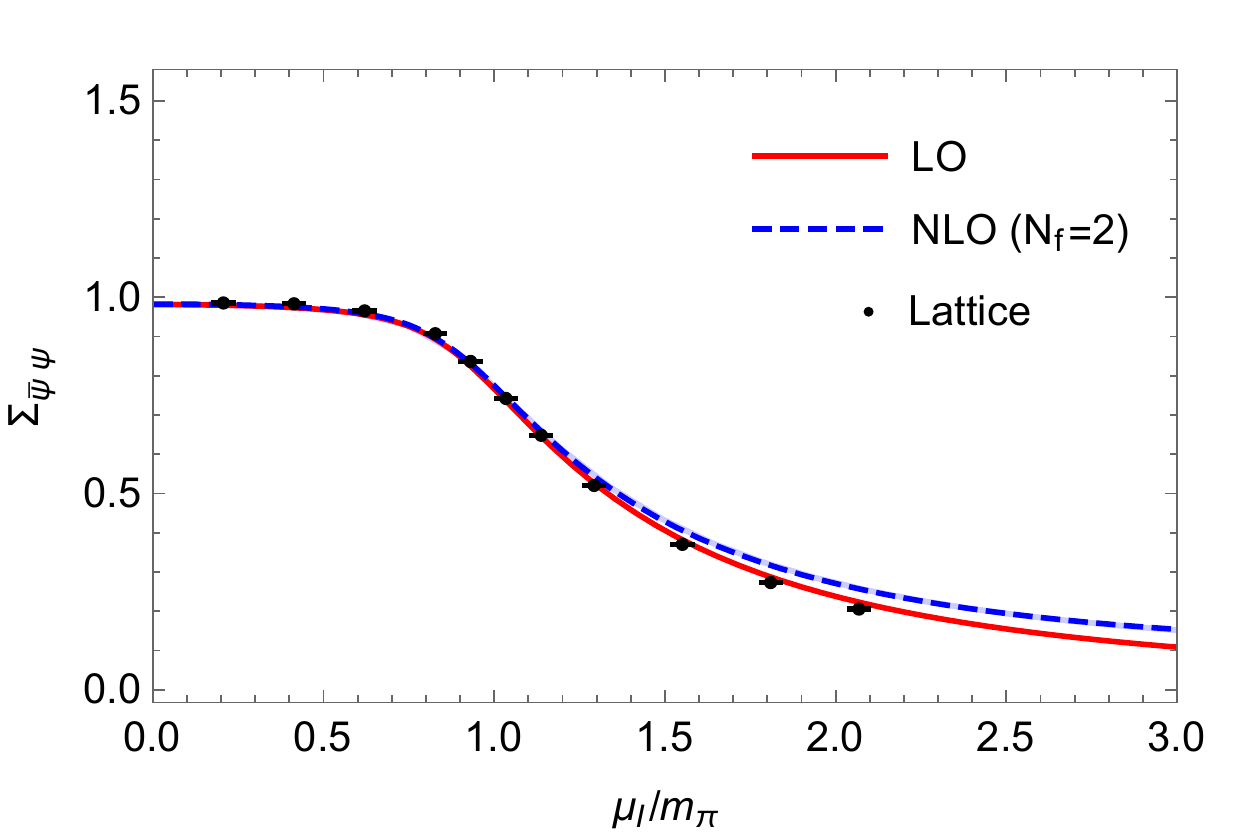}\,\,\,\,\,
\includegraphics[width=0.45\textwidth]{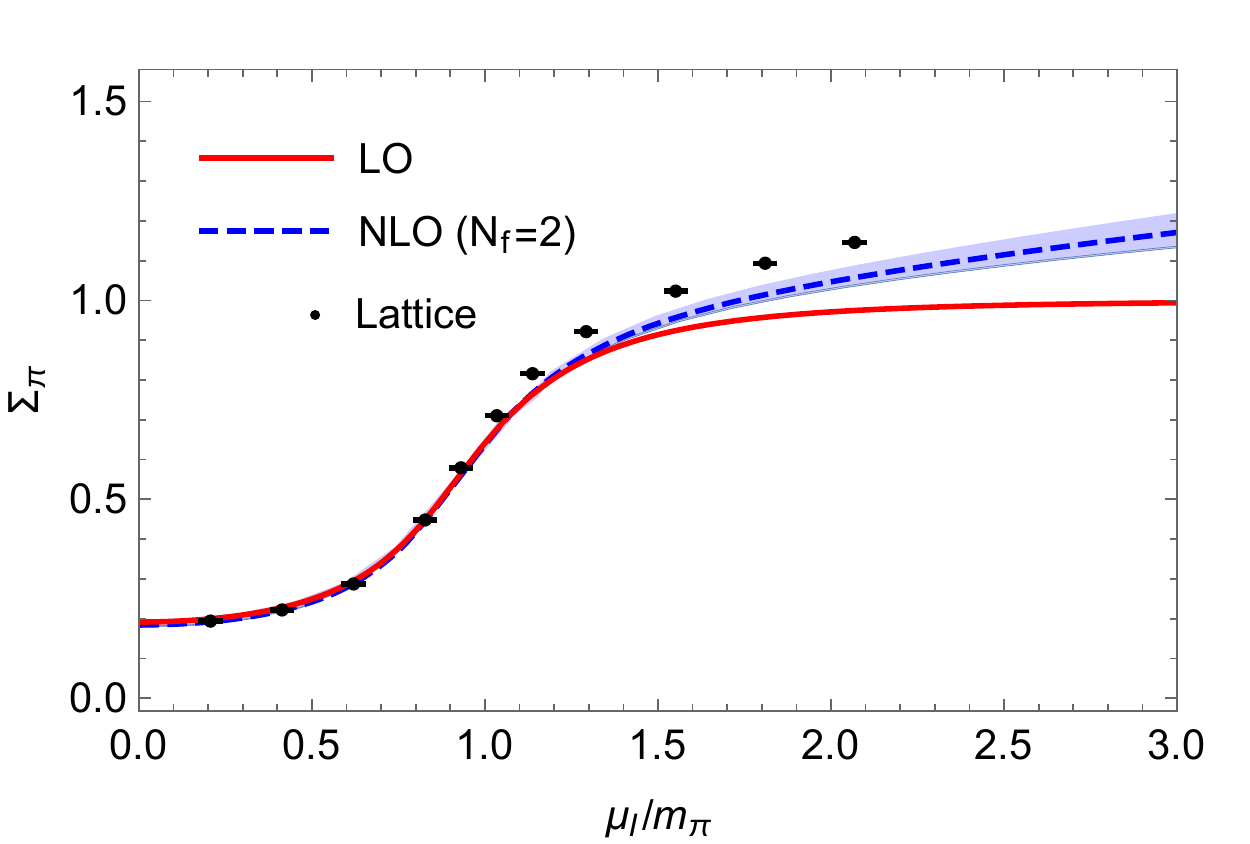}\,\,\,\,\,
\caption{Top: Quark condensate deviation from the normal vacuum value,
  $\Sigma_{\bar{\psi}\psi}$, at $T=0$. Bottom: Pion condensate deviation from the
  normal vacuum value (which is 0), $\Sigma_{\pi}$, at $T=0$ and
  $j=0.00517054m_{\pi}$. See text in~\ref{latcomp}
  for details.}
\label{fig:ccpcj1}       
\end{figure}
In this section, we plot the chiral and pion condensate deviations at $T=0$ with
a non-zero pionic source ($j\neq 0$) and compare our results with lattice
QCD~\cite{latiso,private}. We note that while there is no lattice QCD data
available for comparison at $j=0$, the comparison of finite-$j$ condensate
deviations from $\chi$PT with the lattice allows us to gauge the quality of our
$j=0$ results calculated at next-to-leading order in $\chi$PT. A non-zero $j$
is required to stabilize lattice simulations and consequently $j=0$ results are
``cumbersome" to generate~\cite{gergy3}.

In Fig.~\ref{fig:ccpcj1}, we show the chiral condensate deviation in the top
panel and pion condensate deviation in the bottom panel. The deviations are
calculated at $j=0.00517054 m_{\pi}$, which is the smallest value of the pionic
source for which lattice QCD data is available at $T=0$. In
order to perform this comparison fairly, it is important to know the exact quark
masses in the continuum since this determines the $\chi$PT parameter, $B_{0}$,
on which the condensates depend. As mentioned above, 
continuum quark masses have not been calculated
in the lattice QCD study. 
Consequently, in order to make the comparison quantitative we use the lattice continuum quark masses from a
separate lattice QCD simulation~\cite{BMW} while incorporating uncertainties at the $5\%$ level which are consistent with the uncertainties quoted. We find that condensate deviations are not very sensitive to the quark masses but most sensitive to the uncertainties in the pion mass and pion decay constants.

We also note that due to the presence of an external pionic source,
the ground state explicitly breaks isospin symmetry. Consequently, there is
no second order phase transition as there is in the absence of the pionic
source. Instead, the transition is a crossover involving a range of isospin
chemical potentials within which the chiral and pion
condensates change significantly.

The condensate deviations in Fig.~\ref{fig:ccpcj1} show very good agreement
with the lattice for isospin chemicals potential up to
$\mu_{I}\approx1.5m_{\pi}$. For larger isospin chemical potentials, the lattice
chiral condensate deviation is slightly smaller than the
corresponding deviation from $\chi$PT at next-to-leading order and the lattice pion condensate deviation is moderately larger than the
corresponding deviation from $\chi$PT at next-to-leading order.
For the quark condensate, the LO result is slightly better than the NLO result for
large values of $\mu_I$.
For all values
of the isospin chemical potential, the next-to-leading order
$\chi$PT pion condensate deviation is a significant
improvement over the tree-level results. In particular, there is qualitatively different behavior for the pion condensate deviation at large isospin chemical potential where the deviation does not level off but increases, which is consistent with the behavior of lattice QCD data. The
difference between the tree-level pion condensate deviation and the
corresponding lattice QCD deviation is significantly more
prominent.

\begin{figure}[t!]
\centering
\includegraphics[width=0.45\textwidth]{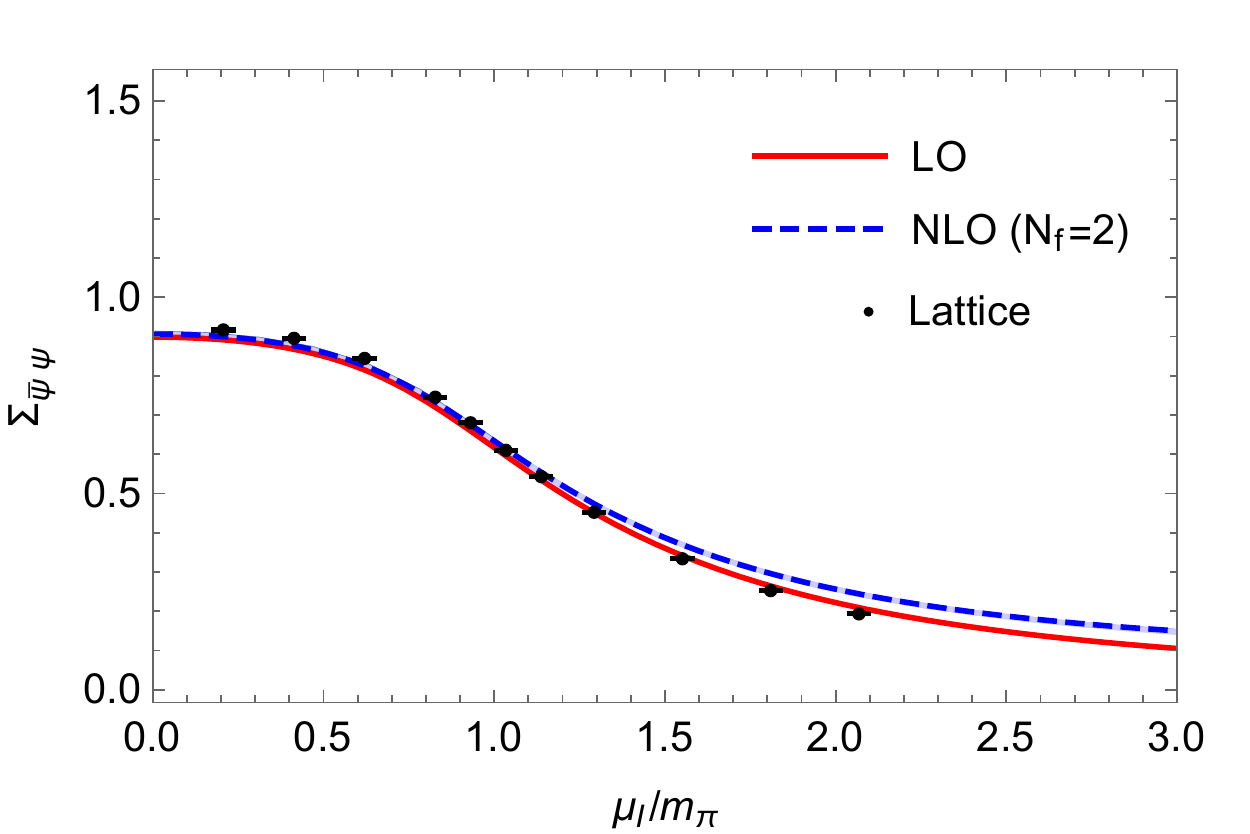}\,\,\,\,\,
\includegraphics[width=0.45\textwidth]{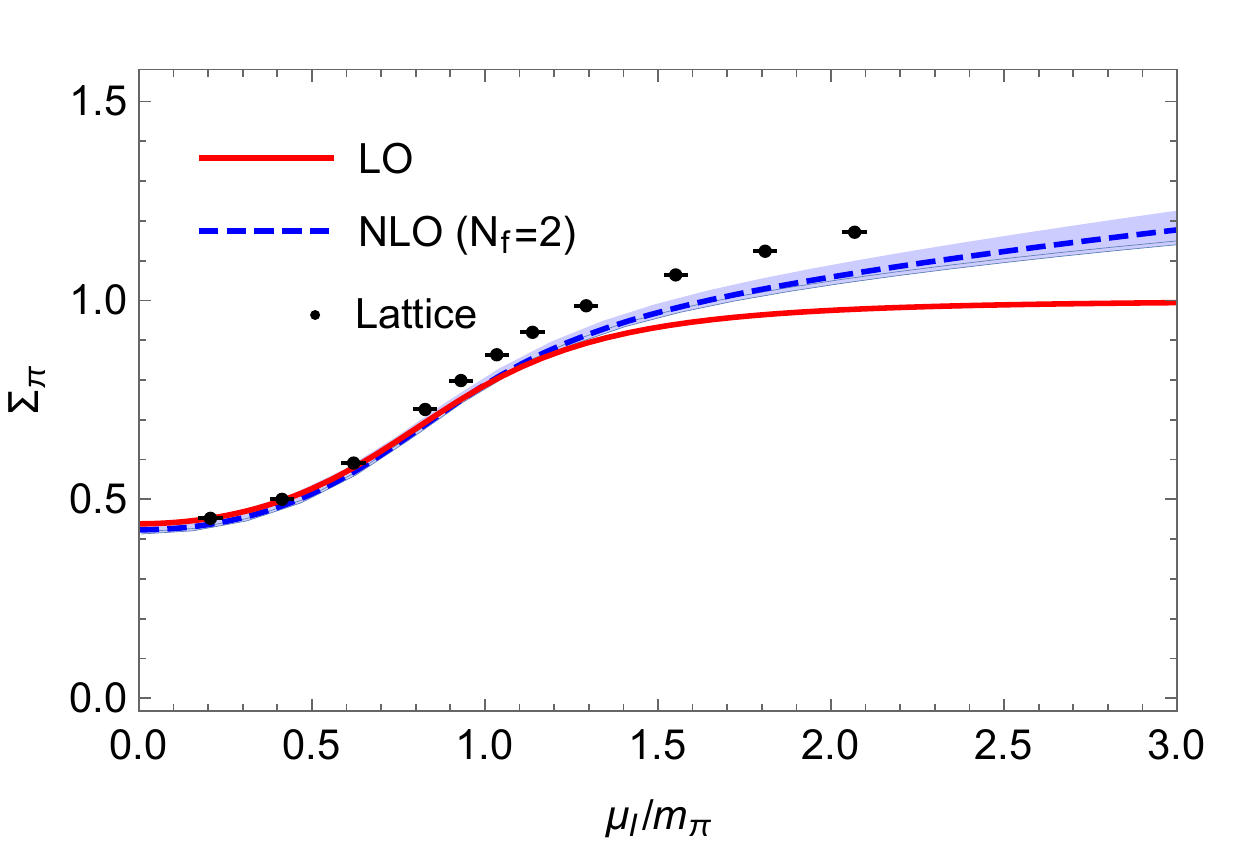}\,\,\,\,\,
\caption{Top: Quark condensate deviation from the normal vacuum value,
  $\Sigma_{\bar{\psi}\psi}$, at $T=0$. Bottom: Pion condensate deviation from the
  normal vacuum value  (which is 0), $\Sigma_{\pi}$, at $T=0$ and
  $j= 0.0129263 m_{\pi}$. See text in~\ref{latcomp}
  for details.}
\label{fig:ccpcj2}       
\end{figure}

Finally, in Fig.~\ref{fig:ccpcj2}, we show the chiral condensate deviation on
the top panel and the pion condensate deviation on the bottom panel for
$j=0.0129263 m_{\pi}$, including $\chi$PT results at tree-level, next-to-leading
order and lattice QCD including uncertainties. As with the previous figure, the
results at next-to-leading order $\chi$PT are an improvement over
tree level deviations except for the chiral condensate
  deviation at large isospin chemical potentials for which there is a mild
  decrease in agreement 
  and for the pion condensate at low isospin chemical potentials, where the decrease in agreement is even milder.
  The improvement is most significant in the pion
condensate deviation, which shows a qualitatively different asymptotic
behavior -- the next-to-leading order pion condensate deviation does not
asymptotically approach 1 as the tree-level result does. The agreement of the deviations
with lattice QCD is very good especially for lower values of isospin chemical
potential consistent with the fact that $\chi$PT is an effective theory with
systematic corrections that increase with the isospin chemical potential. We
also note that the discrepancy between the condensate deviations at larger
isospin chemical potentials is larger for $j=0.0129263 m_{\pi}$ than
$j=0.00517054 m_{\pi}$, which is again consistent with expectations for an
effective theory. This is a general feature
up to the largest values of $j$ used in the simulations.

In conclusion, we have performed a calculation of the quark and pion
condensates at next-to-leading order $\chi$PT in the absence of an external
pionic (pseudoscalar) source for the first time -- the results presented
here can be used to
gauge the quality of future lattice calculation of the chiral and pion
condensate at zero source, a calculation that is currently quite challenging to
perform. We have also calculated the condensates at finite pionic source and
performed a qualitative comparison with the lattice which shows an
improved agreement after we include next-to-leading order corrections.

\section*{Acknowledgements}
The authors would like to thank B. Brandt, G. Endr\H{o}di and S. Schmalzbauer
for providing their lattice condensate data in the presence of a finite pionic
source~\cite{latiso}. The authors would also like to
  acknowledge Martin Mojahed for useful discussions and suggestions.

\end{document}